\documentclass{aa}
\usepackage{graphicx}
\usepackage{longtable}
\usepackage{psfig}
\usepackage{amsmath} 
\usepackage{aas_macros} 
\usepackage{amssymb}
\usepackage{natbib}  
\bibpunct{(}{)}{;}{a}{,}{,} 
\def\ngc#1{NGC$\,$#1} 

\newcommand{\cmd}{color-magnitude diagram}
\newcommand{\cmds}{color-magnitude diagrams}
\def\vmi{\hbox{\it V--I\/}}
\def\bmv{\hbox{\it B--V\/}}
\def\umb{\hbox{\it U--B\/}} 
\def\rmi{\hbox{\it R--I\/}}  
\def\vmr{\hbox{\it V--R\/}}
\def\Deg{${}^\circ$\llap{.}}

\def\deg{${}^\circ$}
\def\min{${}^{\prime}$}

\begin{document}
   \title{Luminosity and Mass Function of the Galactic\\ 
   open cluster \ngc2422 
   \thanks{ Based on observations made at the Cerro Tololo
    Inter-American Observatory, Chile}. }

   \author{L. Prisinzano \inst{1},  G. Micela \inst{2}, S. Sciortino 
   \inst{2}, F. Favata \inst{3} }

   \offprints{loredana@oapa.astropa.unipa.it }

   \institute{ Dipartimento di Scienze Fisiche ed Astronomiche, Universit\`a
di Palermo, Piazza del Parlamento 1, I-90134 Palermo Italy
\and INAF - Osservatorio Astronomico di Palermo, Piazza del Parlamento 1, 90134 
Palermo Italy  \and Astrophysics Division - Space Science Department of ESA, 
ESTEC, Postbus 299, 2200 AG Noordwijk, The Netherlands }

   \date{Received xxx; accepted  xxx}

   \abstract{We present {\em UBVRI} photometry of the open cluster \ngc2422 
   (age $\sim~10^8$ yr) down to a
limiting magnitude $V\simeq19$. These data are used  
to derive the Luminosity
  and Mass Functions   and to study the cluster
spatial distribution. By considering the color-magnitude
diagram data and adopting a representative cluster main sequence, 
we obtained
a list of candidate cluster members based on a photometric criterion. 
 Using a reference field region and an iterative procedure, 
a correction for contaminating field stars has been derived in order 
to obtain the Luminosity and the
Mass Functions  in the
$M=0.4-3.5~M_\odot$ range.
By fitting the
spatial distribution, we infer that 
a non-negligible number of cluster stars lies outside our investigated region. 
We have 
estimated a correction to  the Mass Function of the cluster in order to take 
into account  the "missing" cluster stars. The Present Day Mass
Function of \ngc2422 can be
represented by a power-law of index $\alpha = 3.07 \pm0.08 $ (rms) 
-- the Salpeter Mass Function in this
notation has index $\alpha = 2.35$ -- in the mass range $ 0.9
\leq  M/M_\odot\leq  2.5 $. The index $\alpha$ and the total mass of the cluster
are very similar to those of the Pleiades.\\
}
\titlerunning{\ngc2422}
\authorrunning{Prisinzano et al.} 
%\markboth{Prisinzano et al.}{luminosity function and MF of NGC 2422 }
\maketitle
%
%________________________________________________________________
%%%%%%%%%%%%%%%%%%%%%%%%%%%%%%%%%%%%%%%%%%%%%%%%%%%%%%%%%%%%%%%%%%%%%%%%%%%%%%%%%%%%%%%%%%%%
\section{Introduction\label{intro}}
Accurate determinations of the stellar Initial Mass Function  together
with the star formation rate are fundamental to understand star formation
mechanisms and related astrophysical problems. 
Since Salpeter's estimation of the IMF for stars in the solar-neighborhood
\citep{salp55}, several investigations in Galactic and extra-galactic stellar
systems seem to converge to a universal IMF described by a broken power-law
\citep{scal98}.
Many models of stellar population, chemical evolution and galactic evolution
adopt {\it a priori} a single IMF assuming its universality, although their
results are highly sensitive  to uncertainties in the IMF \citep{kenn98}.

Testing the universality of IMF is a challenge for astrophysicists because
several indirect evidences suggest that the IMF ought to systematically vary 
with the time due to the different star forming conditions \citep{lars98}. 
Nevertheless, no convincing proofs for a variable IMF still exist and evidences
for uniformity of the IMF are deduced by estimates for different populations 
\citep{scal86,scal98,krou02}. 

However, a large  scatter in the logarithmic
power-law index, for stars more massive than $1~M_\odot$, is evident. In order to
understand how large {\it apparent IMF variations} are due to uncertainties 
inherent to any observational estimate of the IMF,
 \citet{krou01} investigated the scatter, introduced by Poisson
noise and dynamical evolution of star clusters, 
of the power-law indices inferred for
 {\it N}-body model populations.
The resultant {\it apparent variation} of the IMF defines a 
"fundamental limit" such that any true
variation in the IMF that is smaller than this fundamental limit is not
detectable. In addition,
determinations of the power-law indices are subject to systematic errors arising
from unresolved binaries.

Being systems of coeval and equidistant stars with the same chemical composition,
open clusters are key samples in investigating the
IMF and its possible spatial and temporal variations.
 Determination of the IMF of open clusters can however be challenging
 because of the
contamination from  background Galactic field stars.
 A further complication comes
from the difficulty to transform the observed Present Day Mass Function (PDMF)
into the IMF using proper assumptions on the stellar and dynamical evolution,
mainly affecting  high and low-mass stars, respectively.

In order to reduce these complications it is convenient to study 
young open clusters or star forming regions.  In 
the low mass star  range  the IMF is   
more uncertain but  stellar evolution effects are not important;
in this mass range the Present Day Mass Function is representative of
the IMF.  

Additionally, for the question of the
universality or variability of the IMF,
it is convenient to compare clusters with same age  
in order to highlight other possible parametric dependence. 
For these reasons we choose to
study the southern Galactic open cluster \ngc2422 which has an age comparable
to the well studied Pleiades cluster.  
  
 The equatorial (J2000.0)
and galactic coordinates  of \ngc2422 are 
RA$= 07^h 36^m 36^s$, Dec=$-$14\deg 30\min~ \citep{lyng87}, 
$l$=230\Deg97, $b$=3\Deg13, respectively.
 Estimates of the fundamental parameters of this cluster, such as   age, 
 distance modulus
and reddening were given in the past
 by various authors as summarized by  \citet{barb02}. 

The age of $\sim 10^8~ $yr was estimated using theoretical isochrones 
\citep{rojo97}; the most recent value  of the  
distance $d=498^{+135}_{-88}$ pc, corresponding to a distance modulus 
$(M-m)_0=8.48^{+0.52}_{-0.42}$,
was deduced by Hipparcos measurements on 4 stars 
\citep{robi99} while the most recent
value of the  reddening $E(\bmv)=0.088$  was reported by
  \citet{damb99}. 

Several papers have been devoted to \ngc2422 in the past: 
the most recent of them report Str\"{o}mgren photometry 
\citep{shob84,niss88,rojo97}, while
the available 
$UBV$ photometric values, either photographic and photoelectric, extend only
down to
 $V \sim 16$ \citep{zug33,lyng59,hoag61,smyt62,sche66,ishm67}.
 Mean photometric data and spectral classification from the former papers
 were compiled by   \citet{merm86} in a catalog  of 212 objects. 
Using this catalog  and 
 564 additional stars from    \citet{ishm67},  
 \citet{barb02} obtained a large literature-based compilation of measured data
 of stars in the field of \ngc2422;  
for some of 
these stars,  X-ray
counterparts were found.

The layout of our paper is  the following. We present, in 
Sect.~2,  our photometry and   astrometry and
in Sect.~3  the method 
adopted to select  the candidate cluster member sample. 
In Sect.~4 we describe
how    the Luminosity and  Mass Functions of \ngc2422 
and the spatial distribution of cluster members were obtained.
 Finally in Sect.~5 we summarize  and discuss our results.
%%%%%%%%%%%%%%%%%%%%%%%%%%%%%%%%%%%%%%%%%%%%%%%%%%%%%%%%%%%%%%%%%%%%%%%%%%%%%%%%%%%%%%%%%%%%
%%%%%%%%%%%%%%%%%%%%%%%%%%%%%%%%%%%%%%%%%%%%%%%%%%%%%%%%%%%%%%%%%%%%%%%%%%%%%%%%%%%%%%%%%%%%
\section{Cluster {\em  UBVRI} Photometry \label{reduc} and Astrometry}
\subsection{Observations  and Data reduction \label{datareduction}}
The data used in this paper are  CCD images in the {\em UBVRI} pass-bands  
 collected at the 0.9-m  telescope of the Cerro 
Tololo Inter-American Observatory (CTIO) on January 30, 1997.  The scale on the 
sky of the instrument is 2.028 arcsec/pixel, for a total field of view
of $1.15\times 1.15$ 
square degrees,    making such an instrument suitable to cover most of the
 apparent size of the cluster.  The
observing log is summarized in Table \ref{obs}. 
While the quality of seeing was limited, all
images were collected 
in photometric conditions.  Possible effects 
on the crowded field photometry, due to  the limited
seeing,  have been ruled out by the artificial
star test described in Sect. \ref{completeness}.  

\begin{table} [htb]
\centering
\tabcolsep 0.3truecm
\caption {Log-book of the \ngc2422 CCD observations}
\begin{tabular}{ ccc} 
\multicolumn{1}{c}{Filter}&  
\multicolumn{1}{c}{Exp. Time} &
\multicolumn{1}{c}{Instrumental seeing} \\  
 & [s] & FWHM[$^{\prime\prime}$] \\
\hline
  {\it U} &   30 &  3.53 \\
  {\it B} &   15 &  3.24 \\
  {\it V} &   15 &  2.70 \\
  {\it R} &   15 &  2.63 \\
  {\it I} &   23 &  3.00 \\
  {\it I} &   10 &  2.99 \\

\hline\\
\end{tabular}
\label{obs}
\end{table}
All the images were pre-processed in a standard way with IRAF, using
the sets of bias and sky flat-field images collected during the observing
night.
The instrumental magnitudes and the  positions of the  stars for each frame
were derived by profile-fitting photometry with the  package DAOPHOT II
and ALLSTAR \citep{stet87}.
Then we used ALLFRAME \citep{stet94} to obtain the simultaneous
PSF-fitting photometry of all the individual frames.
In order to obtain the transformation equations relating the
instrumental  
magnitudes to the standard {\it UBV\/} (Johnson),
$RI$(Kron-Cousins) system, we also derived the instrumental 
profile-fitting photometry 
for the two \citet{land92} fields of
standard stars   SA\-95 and SA\-98 observed during the same night. 
 
In order to obtain the total integrated instrumental magnitudes
we derived aperture photometry for the same stars that we used to
define the PSF, after the digital subtraction of
neighboring objects from the frames.  We used
DAOGROW \citep{stet90} to obtain the aperture growth curves for each
frame and to compute the aperture correction to the profile-fitting photometry.

 The transformation coefficients  to the standard system 
  were derived using
transformation equations of the form: 
\begin{eqnarray}
 v &= V + A_0 + A_1 \times  X + A_2 \times (\vmi), \nonumber\\
 b &= B + B_0 + B_1 \times  X + B_2 \times (\bmv), \nonumber\\
 i &= I + C_0 + C_1 \times  X + C_2 \times (\vmi), \\
 r &= R + D_0 + D_1 \times  X + D_2 \times (\vmr), \nonumber\\
 u &= U + E_0 + E_1 \times  X + E_2 \times (\umb). \nonumber
\end{eqnarray} 
\begin{table}
\centering
\caption {Coefficients of the calibration equations.}
\begin{tabular}{ccc} 
\hline
$ A_0$ & $ A_1 $ & $A_2 $\\
$ 4.637 \pm   0.009$ &  $0.14$  &  $-0.012 \pm   0.010$\\
\hline
$ B_0$ & $ B_1 $ & $B_2 $\\
$ 5.574 \pm   0.002$ &  $0.25$  &  $ 0.095 \pm   0.005 $\\
\hline
$ C_0$ & $ C_1 $ & $C_2 $\\
$ 5.100 \pm   0.002$ &  $0.05$  &  $-0.008 \pm   0.002$\\
\hline
$ D_0$ & $ D_1 $ & $D_2 $\\
$ 4.352 \pm   0.007$ &  $0.10$  &  $-0.029 \pm   0.012$\\
\hline
$ E_0$ & $ E_1 $ & $E_2 $\\
$ 6.989 \pm   0.002$ &  $0.55$  &  $-0.048 \pm   0.008$\\
\hline
\end{tabular}
\label{calib}
\end{table}
\noindent
In these equations $v$, $b$, $i$, $r$ and $u$ are the aperture magnitudes, 
already normalized to 1 sec exposure and $X$ is the airmass.

Due to the  limited number of standard star  observations, we do not have 
a complete coverage of airmass and we have not been able to derive
the extinction coefficients. In order to overcome this problem, 
we used 
typical values for CTIO  available at
http://www.noao.edu/scope/ccdtime/ctio.db; both the adopted  extinction 
coefficients and the best fit values for the zero points
 and the  color coefficients are  summarized in Table \ref{calib}.
Second order color terms were tried and turned out to be negligible in
comparison to their uncertainties.

%% Paragrafo modificato da FF 2002-23-08

 The calibration process we have adopted is based on two
  sequential steps, following a procedure adopted by Stetson (private
  communication).  First, 45 program stars on the \ngc2422 images
  were selected using the condition (following a criterion described in
  \citeauthor{stet93}, 1993, Sect. 4.1) that each star has to be well
  separated from its neighbors, observed in all frames, and with a
  statistic index $\chi$, relative to the goodness of the PSF-fitting
  photometry, less than 1.5. The photometric calibration based on the
  Landolt standards is applied to these selected stars only (which we
  refer to as local standards), which were then used (together with
  the Landolt standard stars) to calibrate the other program stars.
%%%%%%%%%%%%%%%%%%%%%%%%%%%%%%%%%%%%%%%%%%%%%%%%%%%%%%%%%%%%%%%%%%%%%%%%%%%%%%%%%%%%%%%%%%%%%
\subsection{Astrometry} 
The astrometric solution has been computed using 
as reference the recently released  Guide Star catalog, Version 2.2.01
(GSC 2.2). 
%available at http://vizier.u-strasbg.fr/viz-bin/Cat?I/271
At beginning, our pixel coordinate list was matched to the
celestial coordinate list of the GSC 2.2
by  projecting the celestial coordinates onto
a plane and using as reference three stars for which
we had both the pixel and the celestial coordinates from the Hipparcos
catalog \footnote{available at
 http://cdsweb.u-strasbg.fr/cgi-bin/VizieR} \citep{turo93}.
 
 An initial estimate for the linear transformation has been computed 
using the  reference coordinates of the 1350 matched stars. 
Then, a plate solution has been
computed using the same matched pixel and celestial coordinates
by fitting a power series polynomial  (IRAF task {\tt CCMAP}).  
The final accuracy is of 0.24 arcsec. Finally, the IRAF tasks {\tt CCSETWCS}
and {\tt SKYPIX}   were used to obtain the celestial coordinates 
of the total sample. 
%%%%%%%%%%%%%%%%%%%%%%%%%%%%%%%%%%%%%%%%%%%%%%%%%%%%%%%%%%%%%%%%%%%%%%%%%%%%%%%%%%%%%%%
\subsection{The Color-Magnitude Diagram \label{isocmd}}
In order to obtain the color-magnitude diagram (CMD) of the cluster, a 
selection based on the {\tt sharp} parameter was first done following 
\citet{stet87}.
 The {\tt sharp} parameter is related 
to the  angular size of the 
astronomical object allowing to reject 
non-stellar objects as semi-resolved galaxies and 
unrecognized blended double stars, for which {\tt sharp} is significantly
greater than zero, or cosmic rays, for which {\tt sharp} is significantly 
less than zero. 
We considered stellar objects those having the {\tt sharp} parameter in the  
$(-0.2~-~+0.2)$ range. In this way we obtained a list of 36\,101 stellar objects 
for which we have $V$
and $I$ magnitudes. For 35\,732 of these we have {\em VRI} magnitudes,  for 
35\,140 we have {\em BVRI} magnitudes and finally, for 33\,494 we have 
{\em UBVRI} magnitudes.
%stars with UBVRI --> 33494
%stars with BVRI --> 35140
%stars with VRI --> 35732
%stars with VI --> 36101

\begin{figure*}[!ht]
\centerline{\psfig{file=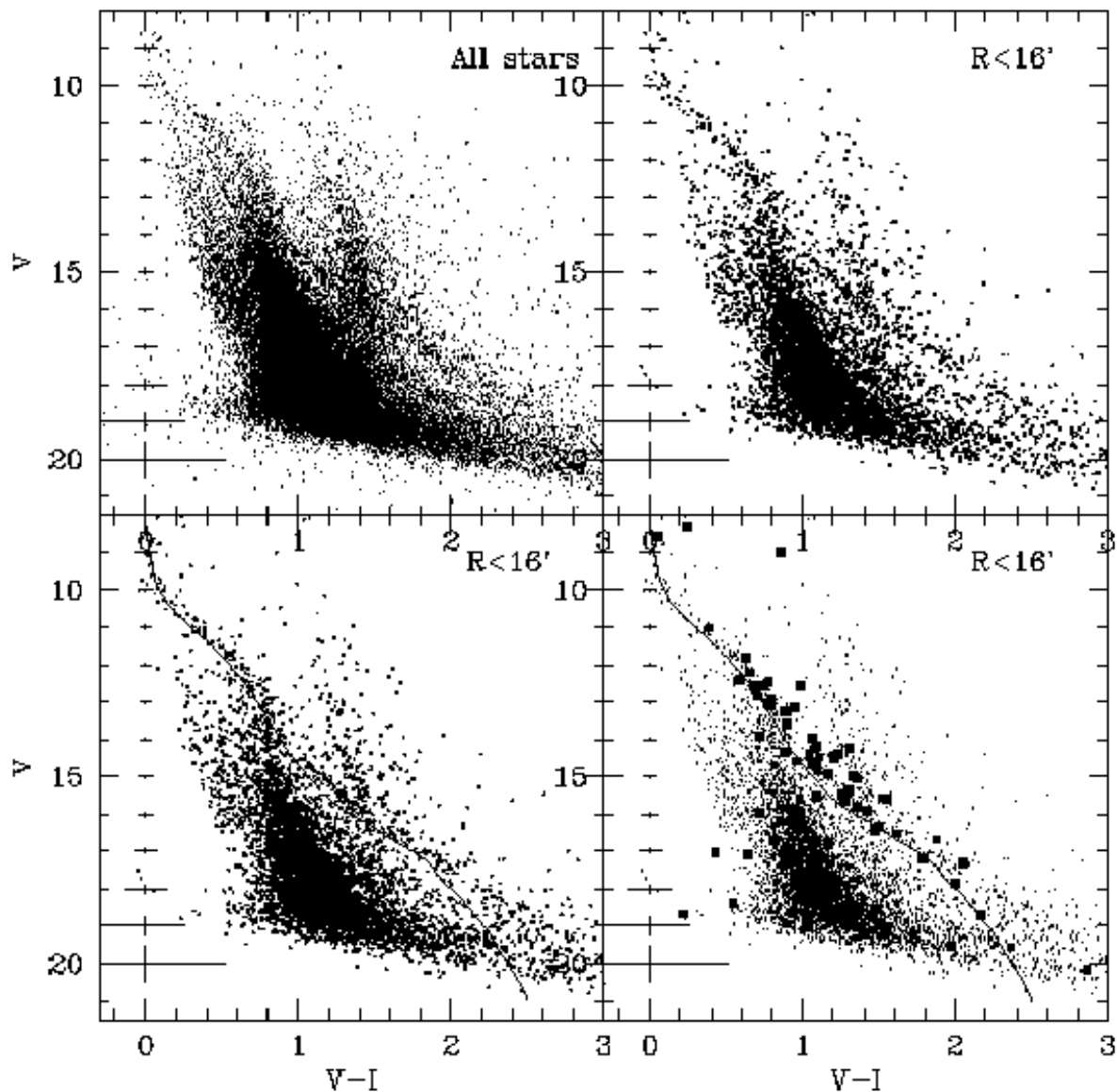,width=17cm,height=17cm}}
\caption{The $V$ vs. \vmi~ \cmd~ for the full catalog ({\it left upper panel})
and for stars within $\sim16$ arcmin from the centroid ({\it right upper
panel} and {\it lower panels}).
The curve is the adopted theoretical isochrone (see text) and the large squares
indicate the positions of the stars which have X-ray counterparts in Barbera
et al. (2002).     
Horizontal bars indicate the median  errors in  color, while vertical bars
indicate the median  errors in  magnitude for bins of one magnitude.}
\label{vi}
\end{figure*}

The $V$ vs. \vmi~  
\cmd~ for all the stellar objects in the \ngc2422 field
is shown in Fig. \ref{vi} ({\it left upper panel}). 
Horizontal bars indicate the median errors in color, while vertical bars
indicate the median errors in  magnitude for bins of one magnitude. 
Clearly, the diagram is
heavily contaminated by background and foreground
stars, as expected from the  position of the cluster in the 
Galactic disk. 

In order to determine the cluster main sequence minimizing
 the contamination effects, we show
in Fig. \ref{vi} ({\it right upper panel}) the $V$ vs. \vmi~ \cmd~ 
approximately corresponding to the cluster core. 
 The centroid of this circular
region of $\simeq 16$ arcmin of radius
has been estimated using the stars with $V \leq  11.5$.

We considered several theoretical isochrones available in literature
\citep{sies00,bara98,gira00,cast99}  
and we have found that the cluster main sequence is well fitted by
the 100 Myr and solar metallicity (Z = 0.02) 
theoretical isochrone computed by 
\citet[ SDF00]{sies00}, for higher mass stars ($M>0.6 M_\odot$), 
and  the one computed 
by  \citet[ BCAH98]{bara98} with a general mixing length parameter
$\alpha=1$, for lower mass stars ($M <0.6 M_\odot$). 
%The isochrone fitting allowed us to validate the 
%distance modulus and the reddening value
%determined by Hipparcos \citep{robi99}  and \citet{damb99},  respectively. 
Both isochrones were transformed into the $V$ vs. \vmi~  plane using
the apparent distance modulus  $(m-M)_V = 8.76$, derived from the distance
modulus  $(m-M)_0 = 8.48$ of  \citet{robi99} and the
interstellar absorption    $A_V= 0.273$, from the standard relation
$R_V=A_V/E(\bmv)=3.1$ \citep{math90} where the reddening $E(\bmv)=0.088$ 
is the value calculated by   \citet{damb99}.   
The color excesses $E(\vmi)=0.11$, $E(\rmi)=0.06$,
$E(\umb)=0.07$  were derived  using   the relations for 
$R_V=3.1$ given by  \citet*{muna96}.
As shown in Fig. \ref{vi} ({\it left lower panel}), the resulting theoretical
isochrone is  in good agreement with the apparent cluster main sequence,
confirming the literature cluster parameters, such as distance
and reddening.

Furthermore, using a set of theoretical isochrones by \citet{sies00} for
different ages, and the well defined main sequence of bright stars in
the $B$ vs.\ \umb~ \cmd, we verified that the \ngc2422 age is $\simeq
10^8$ yr, in agreement with the value given by \citet{rojo97}. 
 
Finally, we matched our photometric data with the X-ray sources 
of \citet{barb02}. The matched stars are plotted with large squares in Fig.
\ref{vi} ({\it right lower panel}). There are two populations of X-ray sources:
a blue faint population, dominated by background objects, and a group of sources
belonging to the cluster main sequence.
The photometric position of the
X-ray  members  allow us to confirm our choice of the main sequence. 

\begin{figure*}[ht]
\centerline{\psfig{file=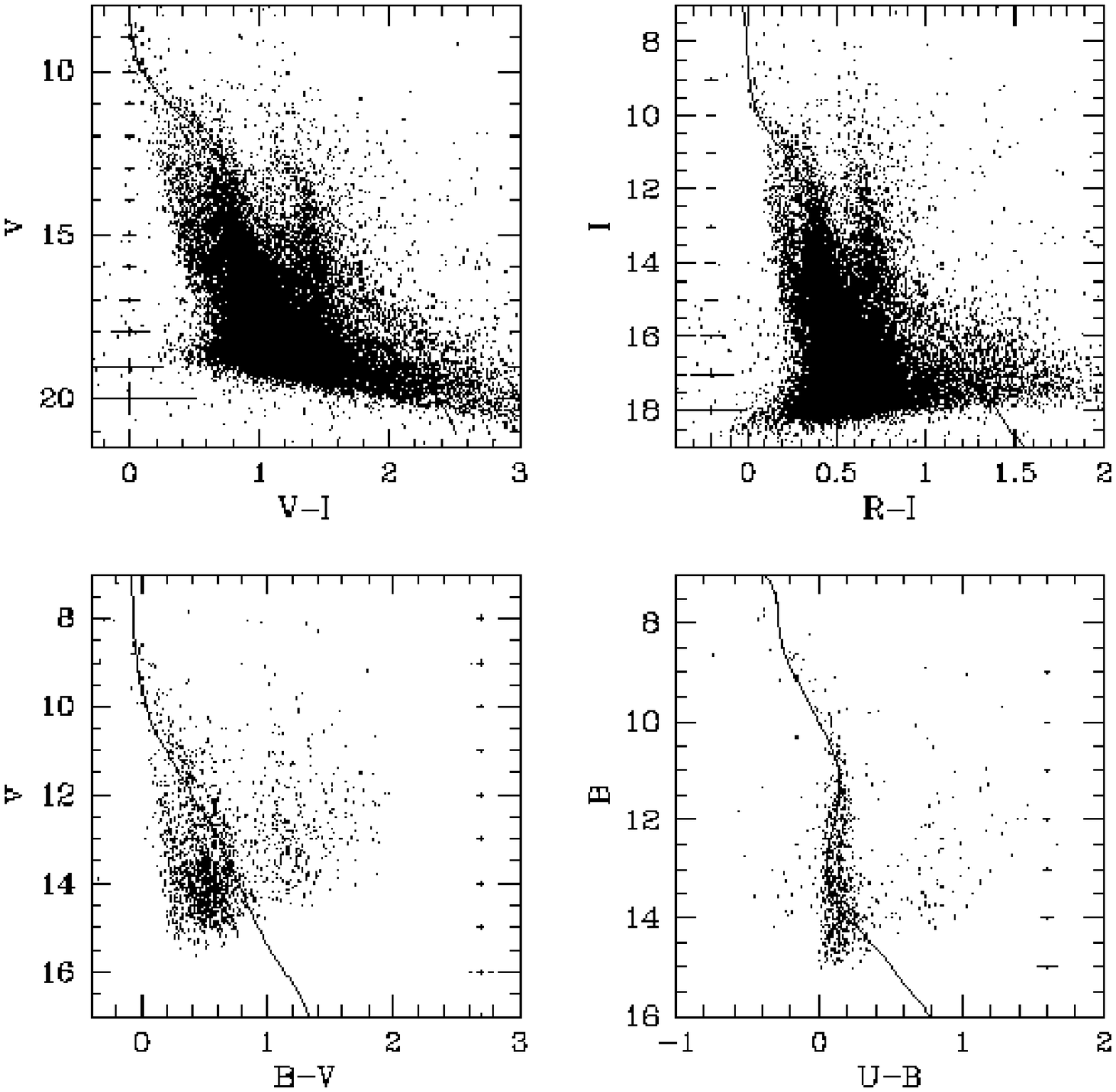,width=17cm}}
\caption{The $V$ vs.  \vmi,  $I$ vs.  \rmi, $V$ vs.  \bmv~ and $B$ vs. 
\umb~ color-magnitude diagrams   for  the stars in the \ngc2422 field. 
 Horizontal bars indicate the median  errors in  color, while vertical bars
indicate the median  errors in  magnitude for bins of one magnitude. The solid
lines are the theoretical isochrones described in the text. In the $V$ vs.  \bmv~ and $B$ vs. 
\umb~ color-magnitude diagrams only stars with $\sigma_B<0.03$ and
$\sigma_U<0.03$, respectively,
are indicated.}
\label{allcmd}
\end{figure*}

In Fig. \ref{allcmd}, we show the  
$V$ vs. \vmi, $I$ vs.  \rmi, $V$ vs.  \bmv~and $B$ vs.  \umb~ \cmds~
for the stars  in the \ngc2422 field.
The curve in the $V$ vs. \vmi~ and in the $I$ vs. \rmi~ \cmds~ 
is the \citet{sies00} theoretical isochrone 
extended to lower stars using the \citet{bara98}
theoretical isochrone as described above.
For the other two diagrams we only considered stars with $\sigma_B<0.03$ and 
$\sigma_U<0.03$, respectively, and 
the \citet{sies00} theoretical isochrone because we use the $V$ vs. \bmv~
and the $B$ vs. \umb~ \cmds~ to select only the bright
cluster members.

%\begin{table} [htb]
%\tabcolsep 0.5truecm
%\caption {Color excess relations for $R_V=3.1$ and for $E(B-V)=0.088$}
%\begin{tabular}{ cc} 
%\hline
%\hline
%$\frac{E(V-I)}{E(B-V)} = 1.25$  &$ E(V-I) = 0.11$\\
%$\frac{E(R-I)}{E(B-V)} = 0.71$  &$ E(R-I) = 0.06$ \\
%$\frac{E(U-B)}{E(B-V)} = 0.82$  &$ E(U-B) = 0.07$ \\
%\hline
%\hline
%\end{tabular}
%\label{redd}
%\end{table}
%%%%%%%%%%%%%%%%%%%%%%%%%%%%%%%%%%%%%%%%%%%%%%%%%%%%%%%%%%%%%%%%%%%%%%%%%%%%%%%%%
%%%%%%%%%%%%%%%%%%%%%%%%%%%%%%%%%%%%%%%%%%%%%%%%%%%%%%%%%%%%%%%%%%%%%%%%%%%%%%%%
%%%%%%%%%%%%%%%%%%%%%%%%%%%%%%%%%%%%%%%%%%%%%%%%%%%%%%%%%%%%%%%%%%%%%%%%%%%%%%%%%
\subsection{Data completeness and photometric errors}
\label{completeness} 
Since we are interested in deriving the Luminosity Function from star counts
and because of the limited seeing of our data,
particular 
attention has been devoted to estimate the accuracy of the photometry and
the completeness of the derived star list.
For this, a list of artificial stars was created and added to the original
frames in order to compare the photometric results of the recovered
artificial stars and the input values. 

In order to avoid overcrowding, the
artificial stars were placed in a spatial grid \citep[cf.][]{piot99} such that   
the separation of the centers 
in each star pair was four PSF radii plus 1 pixel. 
Using a random-number generator
for the $V$ magnitudes, a list of 2470  
 stars ($7\%$ of the total sample) with a flat distribution of the 
instrumental 
$V$ magnitudes between 11 and 21.5 was created. These limits were
chosen on the basis of  the instrumental $V$ vs.  \vmi~ \cmd.  
The   $V$ magnitudes   for each
artificial star  were converted to the other filters using the  fiducial lines 
representing the $V$ vs.  \vmi, $V$ vs.  \bmv, $I$ vs.  \rmi~ and 
$B$ vs.  \umb~ \cmds.

DAOPHOT's {\tt ADDSTAR}
 routine was used to add these artificial stars into copies of the
original data frames, with the appropriate frame-to-frame shifts in their
position and brightness.
Calibrated magnitudes were derived using the same  photometric 
parameters and the same
procedure described in Sect. \ref{datareduction}. 
 
We estimated the completeness fraction as the ratio between the number 
of artificial 
stars recovered simultaneously in the $V$, $I$ and $R$ filters and the number 
of added stars per one magnitude bin. This condition was imposed because the
photometric selection of  low-mass candidate cluster members was based on the
use of
$V$ vs. \vmi~ and $I$ vs.  \rmi~ \cmds.
      
 We found that this ratio is equal to 1 down to $V=18.26$, 
while it decreases to $0.94\pm0.06$ 
  to the limit $V\simeq 19$ of our data. Therefore,
 we can
conclude that our catalog  is complete down to $V=18.26$ over the whole 
$V$ vs.  \vmi~ and $I$ vs.  \rmi~ \cmds~ while it 
is $94\%$ complete to $V=19$, so that 
  a $\sim6\%$ correction  will be required to the number of
faintest stars when computing the 
Luminosity Function. 

Finally, photometric errors were estimated by the differences between
the "observed" magnitudes and colors derived for the artificial stars and their
known input values. We defined the external error 
$\hat{\sigma}_{\rm ext}$ for the $V$ magnitudes and
for the \vmi~ and \rmi~ colors as in \citet*{stet88}; the results
 are summarized
in Table \ref{photoerror}.
%%% See /up10/NGC2422/prisinzano/CROWDING1/photoerror.macro
\begin{table} [htb]
\centering
\tabcolsep 0.2truecm
\caption {External photometric errors from the artificial star
photometry.}
\begin{tabular}{cccc} 
\hline
\hline
$\Delta V$ & $\hat{\sigma}_{{\rm ext},V}$ & $\hat{\sigma}_{{\rm ext},V-I} $ &
$\hat{\sigma}_{{\rm ext},R-I}$  \\
\hline 
 9.5~ - 10.5 & 0.001  &  0.001 &  0.001 \\
 10.5 - 11.5 & 0.001  &  0.003 &  0.003 \\
 11.5 - 12.5 & 0.003  &  0.004 &  0.003 \\
 12.5 - 13.5 & 0.006  &  0.006 &  0.004 \\
 13.5 - 14.5 & 0.007  &  0.009 &  0.007 \\
 14.5 - 15.5 & 0.013  &  0.016 &  0.010 \\
 15.5 - 16.5 & 0.024  &  0.025 &  0.018 \\
 16.5 - 17.5 & 0.043  &  0.058 &  0.039 \\
 17.5 - 18.5 & 0.111  &  0.105 &  0.080 \\
 18.5 - 19.5 & 0.225  &  0.248 &  0.098  \\
\hline\\
\end{tabular}
\label{photoerror}
\end{table}
%%%%%%%%%%%%%%%%%%%%%%%%%%%%%%%%%%%%%%%%%%%%%%%%%%%%%%%%%%%%%%%%%%%%%%%%%%%%
%%%%%%%%%%%%%%%%%%%%%%%%%%%%%%%%%%%%%%%%%%%%%%%%%%%%%%%%%%%%%%%%%%%%%%%%%%%%
\section{Photometric selection of Candidate Cluster Members\label{selec}}
 In order to select possible NGC~2422 cluster members, we used only photometric
informations  as described in the following steps:
\begin{enumerate}
\item{ In the $V$ vs. \vmi~ \cmd, we   selected as   possible candidate
 members  all
those stars which, according to their $\sigma_V$ and $\sigma_{V-I}$ errors, 
belong to a well defined  strip in the CMD.
 The lower envelope of this strip follows 
the representative main sequence for the cluster (see Sect. \ref{isocmd}) 
while the upper envelope is displaced upward by
1 mag   to include binaries. We have choosen 1 mag instead of the canonical
0.75 mag since our isochrone is, in some points of the diagram, slightly
lower than the apparent "true"  sequence of the cluster (see Fig. 1).
With our larger strip we are sure to include X-ray detected stars that are
probable members since significant X-ray emission is a common property of young
stars.
 After this selection, our sample
contains 2059 %a.dat
possible candidate members.}
\item{We considered at first only those stars (1895) %b_1.dat
 for which  {\em UBVRI}~ 
photometry was 
available.  We constructed an analogous strip as above, in the $I$ vs. \rmi~ 
\cmd~ 
and we rejected    from our sample of initial possible candidate
 members  540 %b_1.dat -c_1.dat (1895-1355)
 stars which, according to their $\sigma_I$ and $\sigma_{R-I}$ 
 errors, do not belong to the strip in the $I$ vs. \rmi~ \cmd.}
\item{From the resulting sample, we considered only the 333 %c_up_1.dat
 stars having typical
 error in $B$ magnitude less than 0.03. We saw that this occurs  for $B\leq 16$. 
 Therefore, using the   \citet{sies00} isochrone,  
   we constructed in the $V$ vs. \bmv~ \cmd~ an analogous strip and we 
 rejected those 58 %c_up_1.dat-d_up_1.dat (333-275)
stars  that, according to their $\sigma_V$ and $\sigma_{B-V}$ 
 errors, do not belong to the strip.}
\item{Subsequently, we allowed for the $B$ vs. \umb~ \cmd~ to select the 
bright  stars 
with $U\le13.5$ for which the errors in U magnitude are less than 0.03 mag.
Using the same criterion as above, we defined a strip taking into account
the isochrone's shape 
in this diagram thus further rejecting  %% 275-80 d_up_2.dat-e_up_2.dat
195 stars. 
% The used conditions to defined this strip are in eeup di 
% /up10/NGC2422/prisinzano/NEW/SELEZIONEFOTOMETRICA/fascia4.macro}
\item{Finally, we considered the stars for which only {\em VRI} magnitudes are
available. As above, we considered as possible cluster candidate members
those belonging to the strip defined in the $I$ vs. \rmi~ \cmd.}}
\end{enumerate} 
In our final sample we retained only those stars with $\sigma_I\leq 0.3$ mag.

The selection procedure described above was chosen to take advantage of all
 the available photometric informations    and to avoid
excluding too many possible cluster members.

The photometric/astrometric catalog of the candidate cluster members, 
 containing 1277 stars, is given in Table 4\footnote{available  
in the electronic format via the World Wide Web site 
http://cdsweb.u-strasbg.fr/}
 where we report RA and Dec (J2000.0) coordinates in decimal
degrees, an identification number for each star, the 
$U,~B,~V,~R,~I$ magnitudes and the associated uncertainties. 

\begin{figure*}[!ht]
\centerline{\psfig{file=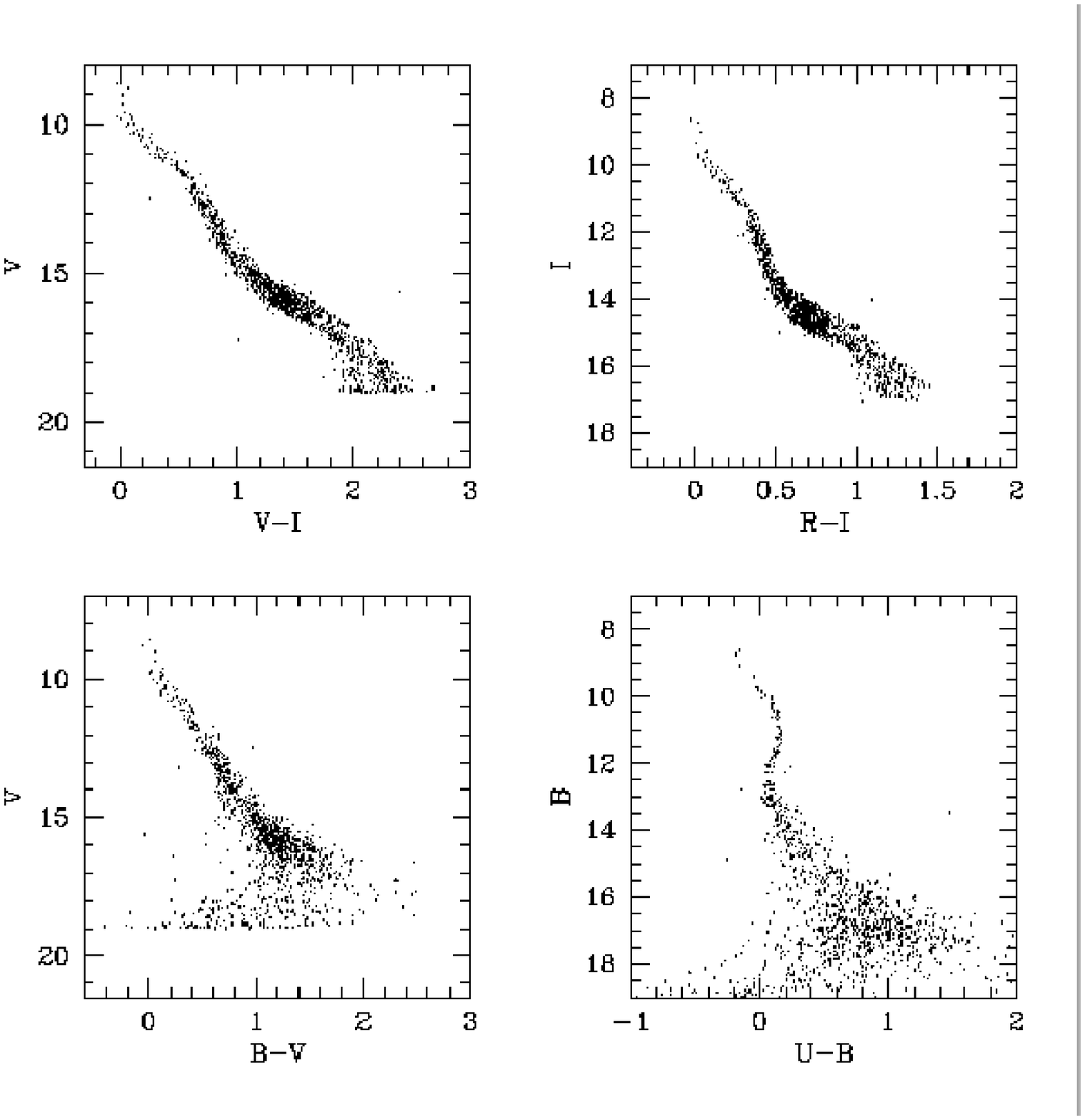,width=17cm}}
\caption{The \cmds~ for the possible photometric 
candidate members of NGC~2422. The sample
has been obtained as described in Sect. \ref{selec}.}
\label{cluster}
\end{figure*}

Fig. \ref{cluster} 
shows these cluster candidates in  the \cmds~ so far considered.  
Clearly, the sample includes contaminating
objects that do not belong to the cluster. The approach to estimate  and to deal
with such a 
contamination will be described in  the following section.
 
%%%%%%%%%%%%%%%%%%%%%%%%%%%%%%%%%%%%%%%%%%%%%%%%%%%%%%%%%%%%%%%%%%%%%%%%%%%%%%%%%%%%%%%%%%%%
%%%%%%%%%%%%%%%%%%%%%%%%%%%%%%%%%%%%%%%%%%%%%%%%%%%%%%%%%%%%%%%%%%%%%%%%%%%%%%%%%%%%%%%%%%%%
\section{Determination of Luminosity and Mass Functions and Spatial Distribution}
\label{sec:corrproc}

%% Modified by FF 2002-08-23

 To construct the Luminosity and Mass Functions of the cluster we
  need to correct the luminosity distribution of our selected sample
  for field star contamination.  To quantify the contamination we
  adopted an iterative procedure schematically summarized in the
  following steps:
  \begin{enumerate}
  \item{We defined as ``field region'' an area of our field of view
      where we expect to find only few cluster stars.  We used such a
      region to quantify the field star contribution and then to
      obtain a first approximation of the cluster Luminosity and Mass
      Functions as well as of the cluster spatial distribution from
      stellar counts in a circular area centered on the cluster
      centroid.  }
    
  \item{By fitting King's empirical profiles \citep{king62} to the
      cluster spatial distribution we then estimated the total number
      of cluster members beyond our cluster region and the number of
      cluster members contaminating the ``field region''.}
    
  \item{After correcting the field star luminosity distribution and
      the field star spatial distribution for cluster stars falling in
      the field region, we re-obtained the cluster Luminosity and Mass
      Functions and the cluster spatial distribution.  By re-fitting
      King's empirical profiles to the new cluster spatial
      distribution we re-estimated the total number of cluster members
      beyond our field of view in order to correct the cluster Mass
      Function.}
    
  \item{We re-corrected the field star luminosity distribution and we
      repeated the above step 3.  The entire procedure was rapidly
      convergent and indeed the ``step 3'' correction was applied only
      three times, since after the third application the previous and
      final corrected Mass Function differ by less than $1\%$, i.e.\ 
      well within the intrinsic uncertainties.}
\end{enumerate}
%\subsection{Contamination by non-members \label{contamin}}
\subsection{First approximation of   Luminosity and Mass Functions 
\label{firsta}}
\begin{figure}[!ht]
\centerline{\psfig{file=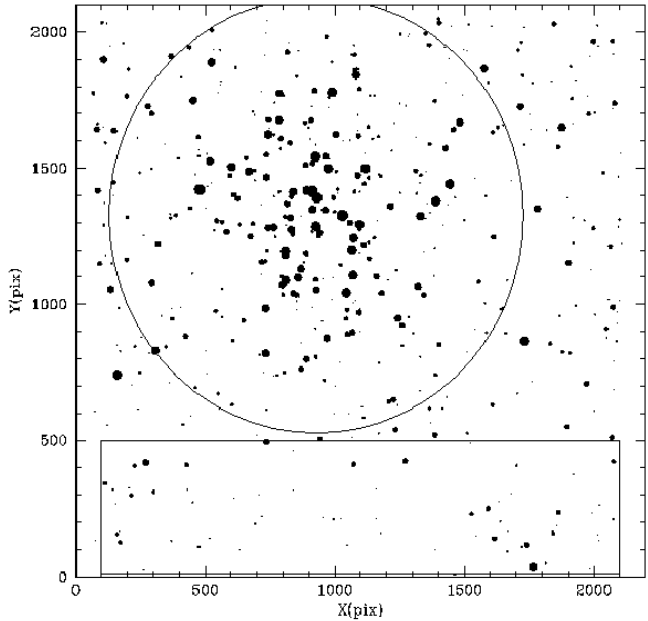,width=9cm,height=9cm}}
\caption{A map of the brightest selected stars in the total field of view
($1.15\times 1.15$ square degrees). Biggest dots  represent
stars with $V\le9$ while smallest dots represent stars with $15.3\le V \le 15.7$.
 The circle delimits the 
 area chosen
to estimate the Luminosity Function of the cluster, while the
rectangle delimits the  area chosen to estimate the contamination rate of
foreground and background stars.}
\label{map}
\end{figure}

  As summarized above, a first approximation to the contamination 
of foreground and background 
sources has been obtained using the lower region of our field where we expect 
 to find only few cluster stars. In fact, as shown in Fig. \ref{map}, 
the cluster is mainly concentrated in the upper region of our field where we
defined the "central cluster region" corresponding to the circular area within the radius
of $27.07$ arcmin.  
A value for the cluster centroid position has been calculated as
the median value of the $X$ and $Y$ coordinates of the bright stars 
($V \leq  12$). This position corresponds to the equatorial coordinates (J2000.0)
$RA=7^h 36^m 34^s.8, Dec=-14^\circ 29^\prime 18^{\prime\prime}.5$.
 We verified that the coordinates of
  the centroid do not change significantly if the $V$ magnitude limit 
changes in the $ 11.5\le V \le 13$ range. 

 We  consider the rectangular region 
in Fig. \ref{map}, of 
$67.6\times 13.52~ {\rm arcmin}^2$ size ($2000 \times 400~ {\rm pixels}$), 
as representative of  the Galactic field (the "field region"), i.e.  
we  suppose that the observed luminosity distribution 
in the rectangular region is given by
\begin{equation}
\hat{N}_{\rm f}(V)=\hat{N}_{{\rm f},0}(V)+\hat{N}_{\rm cl}(V) \simeq \hat{N}_{{\rm f},0}(V),
\label{fieldrec}
\end{equation}
 where $\hat{N}_{{\rm f},0}(V)$ is the "true"
field star  luminosity distribution, while $\hat{N}_{\rm cl}(V)$ is the 
cluster star 
contribution within the rectangular region. 
Therefore,  the field star luminosity distribution in the "central
 cluster region" is given by 
 \begin{equation}
N_{\rm f}(V)=\hat{N}_{\rm f}(V) \frac{A}{\hat{A}},
 \end{equation}
  where 
 $A$  and $\hat{A}$  are  the area of the "central cluster region"
  and of the "field region", respectively. 
 
\begin{figure}[!ht]
\centerline{\psfig{file=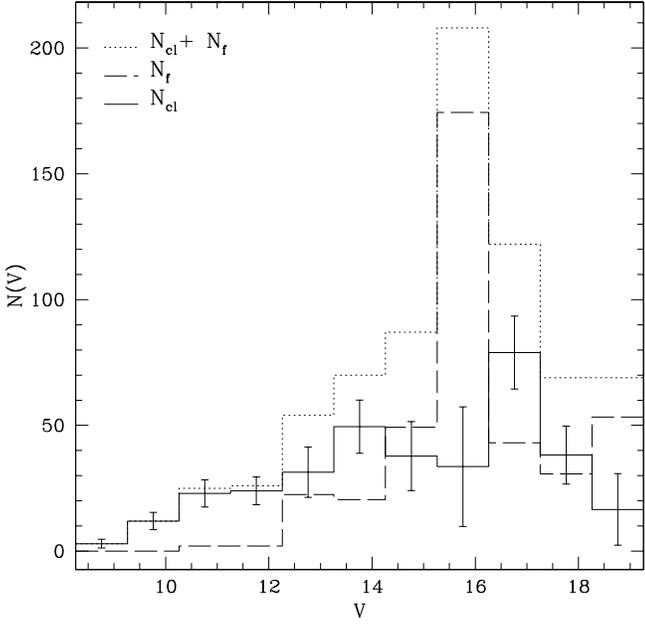,width=9cm,height=9cm}}
\caption{Luminosity distribution of the candidate members  without correction for
 contamination (dotted line), the luminosity distribution
  of the field stars (dashed line) and the
   Luminosity Function of \ngc2422 (solid line)
   obtained as difference between the last two 
  and corrected for incompleteness, as described in Sect. 
  \ref{completeness}.} 
\label{lf1}
\end{figure}

\begin{figure}[!ht]
\centerline{\psfig{file=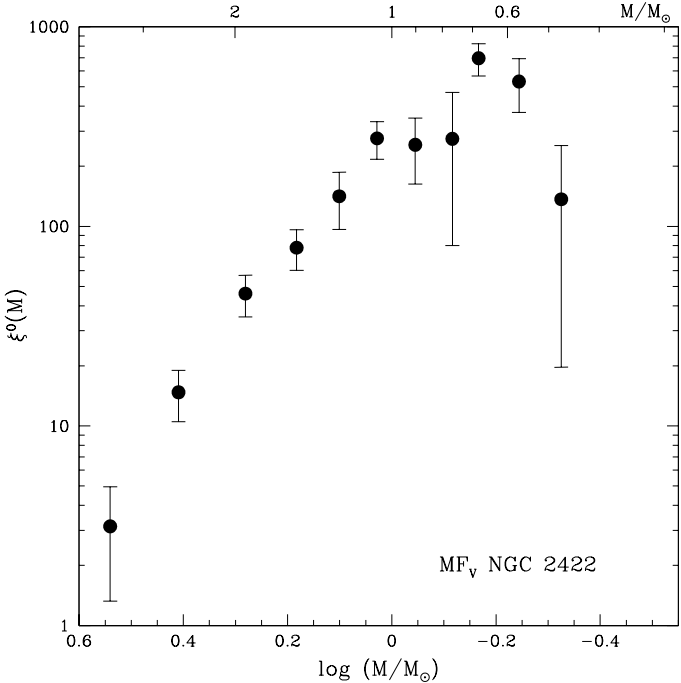,width=9.5cm,height=9.5cm}}
\caption{First approximation of the \ngc2422 Present Day  Mass Function
obtained in our central cluster region and without any correction.
$\xi^0(M)$ values are given in number per logarithmic mass
 unit.}
\label{mf}
\end{figure}

The \ngc2422 Luminosity Function, $N_{\rm cl}(V)$, has hence been 
obtained subtracting the field 
star luminosity distribution, $N_{\rm f}(V)$, from the luminosity distribution 
of  all the stars in our
sample within the "central cluster region", $N_{\rm t}(V)$. 
 Fig. \ref{lf1} shows the contaminated luminosity
 distribution  of the
 candidate members 
  (dotted line), the luminosity distribution of the field stars 
 (dashed line) and the
Luminosity Function of \ngc2422 obtained as described above  and 
corrected for incompleteness as derived in Sect. \ref{completeness} 
(solid line).  
We found that the contamination rate from field stars 
is negligible for the brightest stars, but 
starts to become significant at $V \simeq 13$ where it is of the order of the 
$40\%$. At fainter magnitudes this value increases with a peak of the
order of $\sim90\%$ around
$V\simeq 16$ where the contribution of the field giant stars is dominant.
The total number of the  cluster members estimated at this step is 347.   
 %%%%%%%%%%%%%%%%%%%%%%%%%%%%%%%%%%%%%%%%%%%%%%%%%%%%%%%%%%%%%%%%%%%%%%%%%%%%%%%%%
%%%%%%%%%%%%%%%%%%%%%%%%%%%%%%%%%%%%%%%%%%%%%%%%%%%%%%%%%%%%%%%%%%%%%%%%%%%%%%%%% 
  The Luminosity Function obtained at this step 
has been transformed into a first approximation of 
the Mass Function, $\xi^0(M)$,  using the
 mass-visual magnitude relation  derived from the same 
models that we used to fit the cluster \cmds. The resultant 
first approximation of the Present Day Mass
Function, extending from
$0.5$ to $3.5~M_\odot$, is shown in Fig. \ref{mf}. 

As already mentioned, due to dynamical evolution,  
the angular size of the   cluster is almost certainly
larger than the investigated "central cluster region" hence
this Mass Function is not representative of the whole cluster mass 
distribution.  In order to  estimate the correction to the Mass Function
due to the dynamical evolution, 
we studied the spatial
distribution of \ngc2422 as described in the following Sections.
 
%%%%%%%%%%%%%%%%%%%%%%%%%%%%%%%%%%%%%%%%%%%%%%%%%%%%%%%%%%%%%%%%%%%%%%%%%%%%%%%%
\subsection{First approximation of the Spatial Distribution}
\label{spatial}
The spatial distribution of the cluster members
was  studied in order  to investigate
the cluster dynamical evolution and to determine the cluster dynamical
parameters. As first approximation, we corrected the radial surface density
distribution of candidate cluster members  for foreground
and background contaminating stars assuming that the field 
stars are uniformly  distributed, i.e. 
\begin{equation}
\frac{dN}{dr}=2 \rho \pi r,
\end{equation}
where $r$ is the distance from the cluster center and $\rho$ is the 
star distribution for unit of area in the rectangular region  considered 
in Sect. \ref{firsta}.

\begin{figure}[!ht]
\centerline{\psfig{file=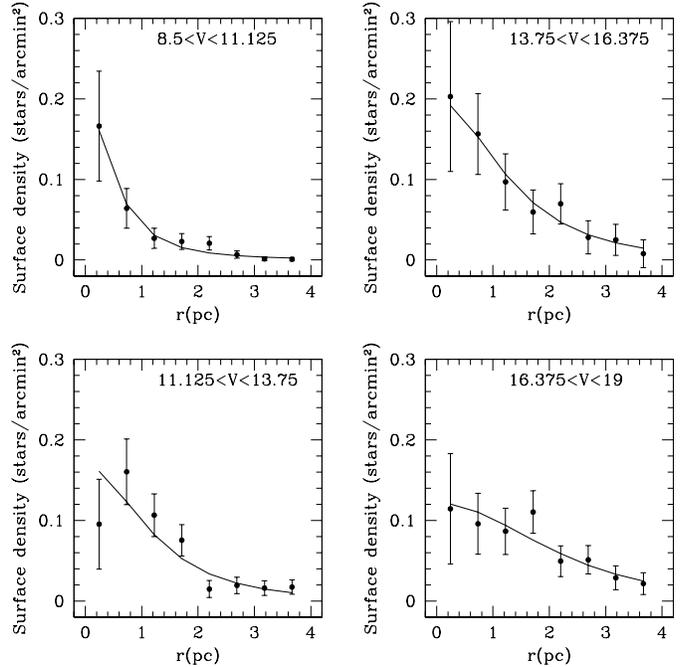,width=9.5cm,height=9.5cm}}
\caption{The radial distribution of stellar density in \ngc2422, for 4 different
bins of magnitude. }
\label{radprof}
\end{figure}

 In order to  take
into account the dynamical evolution and mass segregation effects due to the
energy equipartition (see  \citeauthor{krou01}, 2001 and relative references),
we  subdivided the  stars into 4 bins of $V$ magnitude as in Table 
\ref{radist}.
In Fig. \ref{radprof} we present the cluster surface density profiles  in
${\rm stars/arcmin}^2$ as a function
of radius in   parsec, assuming the distance value of $497.5$ pc
by   \citet{robi99}.  
%%%%%%%%%%%%%%%%%%%%%%%%%%%%%%%%%%%%%%%%%%%%%%%%%%%%%%%%%%%%%%%%%%%%%%
%%%%%%%%%%%%%%%%%%%%%%%%%%%%%%%%%%%%%%%%%%%%%%%%%%%%%%%%%%%%%%%%%%%%%%%%
We  fitted 
King's empirical profiles \citep{king62}, given by
\begin{equation}
\label{eq_king}
\rho (r)=\rho_0 (\frac{1}{\sqrt{1+(r/r_{\rm c})^2}}-\frac{1}{\sqrt{1+(r_{\rm t}/r_{\rm c})^2}})^2,
\end{equation}  
to   the surface density profiles.      
In this equation, $\rho_0$ is a normalization constant, $r_{\rm c}$ is the "core
radius" determined by the internal energy of the system 
%, that is the radius where the surface density falls to half 
%its central value,  
and $r_{\rm t}$ is the tidal radius where the cluster disappears. 

The tidal radius was  calculated as in  \citet{jeff01}
as $r_{\rm t}=1.46~ M_{\rm c}^{1/3}$, where $M_{\rm c}$ is the cluster
 mass in 
solar masses. 
In order to estimate  the total mass of the cluster we integrated 
the Mass Function obtained in Sect. \ref{firsta} 
and  found a value of $ M_{\rm c}=357\pm 35~ M_\odot$, corresponding to the 
tidal radius value $r_{\rm t}=10.36\pm0.34$ pc. 

The results  of King's empirical profile 
fitting, with the 
1-$\sigma$ uncertainty estimates for the  parameters,
are reported in Table \ref{radist}   
and are shown in Fig. \ref{radprof}.
Comparing  
the $\chi^2$ values with  the number of degrees of freedom ($\nu=6$),  
we  conclude
that the fits are all acceptable.
%%%%%%%%%%%%%%%%%%%%%%%%%%%%%%%%%%%%%%%%%%%%%%%%%%%%%%%%%%%%%%%%%%%%%%%%%%%%%%%%%%%%%%%%%
\begin{table*} [htb]
\addtocounter{table}{+1}
\centering
\tabcolsep 0.3truecm
\caption { Results of King's empirical profile fitting to the radial surface
density distribution. Columns 1 and 2
give the  $V$ magnitude ranges and the corresponding mass
ranges; Col. 3 the fitted normalization constant, Col. 4 the assumed
tidal radius, Col. 5 the fitted core radius, Col. 6  the $\chi^2$ of the
fitting. Finally, Col. 7  lists the number of counted cluster members
in our investigated
region for each magnitude subset, while Col. 8 and 9 are the total number of
cluster members predicted by the fitted model
within the investigated radius ($\sim 3.7$ parsec) and 
the tidal radius, respectively. 
The values are given for each of the four step of the adopted iterative procedure.}
\begin{tabular}{ccccccccc}%} 
\hline
\hline
$\Delta V$ & $M$&$\rho_0  $&$r_{\rm t}$&$r_{\rm c}$ & $\chi^2$ &$N_{\rm sur}^{\rm cl} $&
$N_{\rm sur}^{\rm cal} $& $N_{\rm tid}$\\  
&$(M_\odot)$&$({\rm stars/arcmin}^2)$&(pc)&(pc)&&&&\\
\hline
\hline
%&&&&&&&& \\
&&&ITERATION 0&&&&&\\
%&&&&&&&& \\
\hline
~8.50 - 11.13  &3.7 - 1.7& $0.21 \pm 0.10 $ &$10.36$&$ 0.59  \pm 0.22 $  & 4.2 & 34& 29 &33 \\
11.13 - 13.75  &1.7 - 1.0& $0.22 \pm 0.13 $ &$10.36$&$ 1.34  \pm 0.46 $  & 8.5 & 88& 76 &95 \\
13.75 - 16.38  &1.0 - 0.7& $0.27 \pm 0.24 $ &$10.36$&$ 1.49  \pm 0.73 $  & 1.3 &106&101 &129\\
16.38 - 19.00  &0.7 - 0.4& $0.22 \pm 0.26 $ &$10.36$&$ 2.66  \pm 1.48 $  & 2.5 &119&111 &164\\
%&&&&&&&& \\
%\hline
\hline
&&&ITERATION 1&&&&&\\
%&&&&&&&& \\
\hline
~8.50 - 11.13  &3.7 - 1.7& $0.21 \pm 0.10 $ &$10.57$&$ 0.62  \pm 0.23 $ & 4.0& 35& 30& 34 \\
11.13 - 13.75  &1.7 - 1.0& $0.23 \pm 0.13 $ &$10.57$&$ 1.36  \pm 0.44 $ & 8.7& 93& 81&103 \\
13.75 - 16.38  &1.0 - 0.7& $0.27 \pm 0.24 $ &$10.57$&$ 1.55  \pm 0.74 $ & 1.3&113&108&140 \\
16.38 - 19.00  &0.7 - 0.4& $0.24 \pm 0.25 $ &$10.57$&$ 2.65  \pm 1.30 $ & 2.5&134&124&185 \\
%&&&&&&&& \\
%\hline
\hline
&&&ITERATION 2&&&&& \\
\hline
~8.50 - 11.13  &3.7 - 1.7&$0.21 \pm 0.10 $ &$10.60$&$0.62  \pm 0.23 $  & 3.9& 35& 30&35\\
11.13 - 13.75  &1.7 - 1.0&$0.22 \pm 0.12 $ &$10.60$&$1.40  \pm 0.45 $  & 8.7& 94& 80&102\\
13.75 - 16.38  &1.0 - 0.7&$0.27 \pm 0.24 $ &$10.60$&$1.58  \pm 0.76 $  & 1.3&115&109&142\\
16.38 - 19.00  &0.7 - 0.4&$0.24 \pm 0.26 $ &$10.60$&$2.78  \pm 1.41 $  & 2.4&136&125&191 \\
%&&&&&&&& \\ 
%\vspace{0.5mm}\\
\hline
&&&ITERATION 3&&&&& \\
\hline
~8.50 - 11.13  &3.7 - 1.7&$0.21 \pm 0.10 $ &$10.61$&$0.62  \pm 0.23 $  & 3.9& 35& 30&35\\
11.13 - 13.75  &1.7 - 1.0&$0.22 \pm 0.12 $ &$10.61$&$1.39  \pm 0.47 $  & 8.7& 94& 80&102\\
13.75 - 16.38  &1.0 - 0.7&$0.27 \pm 0.24 $ &$10.61$&$1.58  \pm 0.75 $  & 1.3&115&109&142\\
16.38 - 19.00  &0.7 - 0.4&$0.24 \pm 0.27 $ &$10.61$&$2.86  \pm 1.47 $  & 2.4&137&127&195 \\
\hline
\hline
\end{tabular}
\label{radist}
\end{table*}
%%%%%%%%%%%%%%%%%%%%%%%%%%%%%%%%%%%%%%%%%%%%%%%%%%%%%%%%%%%%%%%%%%%%%%%%%%%%%%%%%%%%%%%%%%
From the resulting distribution we can see that, while the distribution 
of the brightest stars
vanishes to zero as the radius increases 
(at the top on the left of Fig. \ref{radprof}),  
the other profiles tend to non negligible  values when fainter
  $V$ magnitudes are considered.  
This result suggests an evidence of mass segregation
of high mass cluster stars toward the center of the cluster and   
low mass cluster stars 
out of our investigation region of $27 $ arcmin of
 radius. 
 This conclusion is confirmed by the increase of the core radius
 as the mass decreases. 
 
 In order to estimate the fraction of low-mass stars beyond
 the "central cluster region",
  we extrapolated the surface density profiles as far as the
 tidal radius. Using the integral of  equation
\ref{eq_king} as given by \citet{king62}, 
we calculated, for each   bin of magnitude, 
 the total number of cluster stars within the tidal
 radius, indicated in Table \ref{radist} by 
 $N_{\rm tid}$.
We compared the calculated total number $N_{\rm tid}$
with the number of cluster members $N_{\rm sur}^{\rm cl}$,
detected in our survey, that is within  $r_{\rm sur}=3.7$ pc 
and we found that  while the stars with $V \lesssim 11$ are all in the investigated region,
the fraction of the cluster
stars lying outside our survey is 
$8\%$ for $11.13 \leq  V \leq  13.75$,
$22\%$ for $13.75 \leq V \leq  16.38$ and
$38\%$ for $16.38\leq V\leq 19.00$. In order to verify the consistency of these
numbers we also integrated King's empirical profile within the radius 
$r_{\rm sur}$, indicated by $N_{\rm sur}^{\rm cal}$,
finding consistent values with the number of the cluster stars found in the
central cluster region ($N_{\rm sur}^{\rm cl}$). 

 The results of the King's empirical profile integration also suggest that 
the cluster
star contribution within the adopted "field region", $\hat{N}_{\rm cl}(V)$  
is non-negligible and that the
estimated field star luminosity and spatial distributions have to be corrected for the 
cluster star contamination. In order to estimate the correction factor, we split the
"field region" in a grid of $5 \times 20$ subregions and we calculated 
the King's profile 
surface density, $\rho(r_i)$, where $r_i$ is 
the radius of the centers of each subregion $i$. Therefore, the total
number of cluster stars within the "field region", per  magnitude bin, is
given by 
\begin{equation}
\hat{N}_{\rm cl}(V)=A_{\rm s}\sum_i{\rho(r_i)},
\end{equation}
where $A_{\rm s}$ is the area of each subregion.

Finally, using equation (\ref{fieldrec}), we estimated a more
accurate field star distribution as
\begin{equation}
\hat{N}_{{\rm f},0}(V)=\hat{N}_{\rm f}(V)-A_{\rm s}\sum_i{\rho(r_i).}
\label{fieldcor}
\end{equation}
%%%%%%%%%%%%%%%%%%%%%%%%%%%%%%%%%%%%%%%%%%%%%%%%%%%%%%%%%%%%%%%%%%%%%%%%%%%%%%%%%%%%%%
\subsection{Luminosity and Mass Functions}

%% paragraph modified by FF 2002-08-23

 Using the value of $\hat{N}_{{\rm f},0}(V)$ from
  Eq.~(\ref{fieldcor}) and the iterative approach
  described in Sect.~\ref{sec:corrproc}, we recomputed the Luminosity Function
  of the cluster taking into account the presence of cluster stars in the
  ``field region''.
  
  This final Luminosity Function has been converted into a Mass
  Function that has been integrated to compute a new value for the
  total mass ($M_{\rm c}\simeq385~ M_\odot$), corresponding to $r_{\rm
    t}\simeq10.61$ pc. Dynamical parameters were also obtained in the
  various iterations, and by integration of King's profiles we
  interpolated the ratio $R=N_{\rm sur}^{\rm cl}/N_{\rm tid}$,
  deriving a factor $1/R$ for the Mass Function to correct for the
  cluster stars lying beyond the investigated cluster region.  The
  corrected Mass Function has been computed by the expression:
\begin{equation}
\xi^1_{c}(M)=\xi^1(M)/R.
\end{equation} 

%We repeated the above step until  the requirement
%\begin{equation}
%\frac{\xi^n_{c}(M)-\xi^{n-1}_c(M)}{\xi^{n-1}_c(M)}\le0.01
%\end{equation} 
%was satisfied in order to stop the procedure.
%In our case, this condition was verified at $n=3$.

\begin{figure}[!ht]
\centerline{\psfig{file=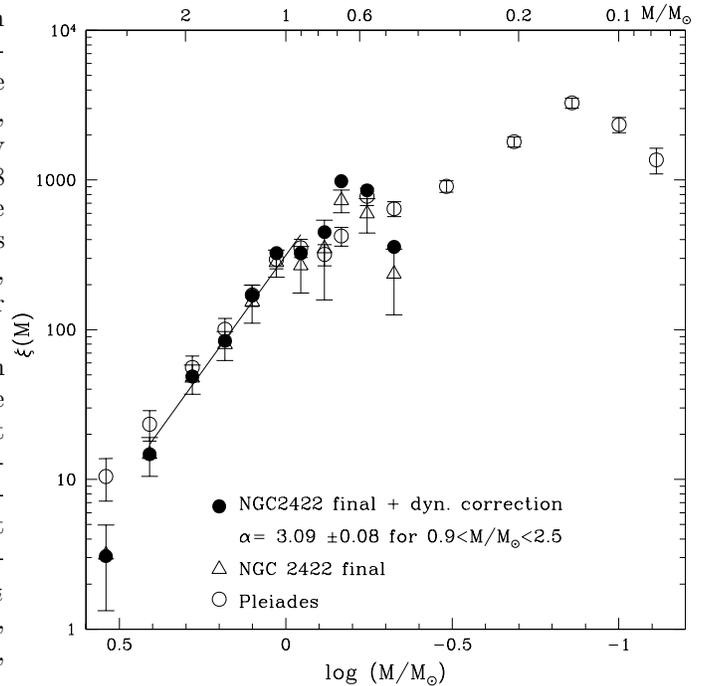,width=9.5cm,height=9.5cm}}
\caption{Comparison of the \ngc2422 
Mass Function
 corrected for dynamical evolution effects (filled points) 
with the \ngc2422 Mass
Function without correction for dynamical evolution (empty triangles) 
and with the Pleiades one (empty circles). The line is the power law fitting
the MF in the $0.9~-~2.5~M_\odot$ range. 
$\xi(M)$ values are given in number per  logarithmic mass unit. }
\label{mfcorr}
\end{figure}

In Fig. \ref{mfcorr}   we compare the final Mass Function, 
$\xi_{c}(M)$, with the Mass Function  
derived  from the central cluster region but without correction
for dynamical evolution. We also compare them
with  the Pleiades Mass Function  derived using the   \citet*{lee95} 
Luminosity Function    and the same
mass-luminosity relation  used for \ngc2422 in the present work.

   According  to the recently proposed analytical IMF forms \citep{krou02},
we  considered a 
multi-part power-law IMF to fit our data. 
We found that  in the mass range 
$0.9\leq M/M_\odot\leq  2.5$  
%$dN/dlogM \propto M^{-\alpha}$ with 
the value of $\alpha$ is 
$3.09\pm0.09$ (rms) and $2.63 \pm 0.16$ for \ngc2422 and
 Pleiades, respectively.
The latest value is very close to the value 2.67  derived by  
\citet{barr01} for the Pleiades,
using the  \citet{hamb99} data in the $0.6~-~10~M_\odot$
mass range.
The two values here derived 
are in  better agreement, within the errors, with  the   \citet{scal98} value
$\alpha =2.7\pm0.3$ rather than with the   \citet{salp55} value
  $\alpha = 2.35$.
  
%Fig. \ref{mfcorr} 
%also shows that dynamical corrections are required to obtain
%a correct cluster Mass Function. It is worth noting that the inclusion of 
%dynamical  corrections is essential to obtain a \ngc2422 Mass Function that
%results similar to that of the Pleiades.
  
Using the final Mass Function, we also estimated the corrected cluster
total mass ($M_{\rm c,cor}\simeq453~ M_\odot$)
and we compared this value 
with the 
\ngc2422 total mass estimated within the "central cluster region",  
$M_{\rm c}\simeq385~ M_\odot$. 
In the same 
mass range ($0.5~-~3.5~M_\odot$), the
total mass of the Pleiades   is $M_{\rm Pl}=467\pm 26~M_\odot$, 
a value very similar to the one of \ngc2422.  
 We note that our estimate of the total mass of NGC\,2422 is, in any case,
a lower limit, since we do not have taken into account the presence of 
companions in photometric binaries.

%%%%%%%%%%%%%%%%%%%%%%%%%%%%%%%%%%%%%%%%%%%%%%%%%%%%%%%%%%%%%%%%%%%%%%%%%%%%%%%%%%%%%
%%%%%%%%%%%%%%%%%%%%%%%%%%%%%%%%%%%%%%%%%%%%%%%%%%%%%%%%%%%%%%%%%%%%%%%%%%%%%%%%%%%%
\section{Summary and Conclusions}
Using  {\em UBVRI} images, covering a field of view of
 $1.15\times 1.15$ square degrees, we extracted a
photometric and astrometric catalog  of the stars in the field of \ngc2422 down to $V\simeq19$ 
($M\simeq 0.4~M_\odot$). 
Adopting a representative cluster main sequence to the
%(Section \ref{isocmd})
\cmds, a photometric criterion was defined to obtain
a list of candidate cluster members.  
A test with artificial stars allowed us to verify that our data are 
complete down to
$V=18.26$ while a correction of $\simeq 6\%$ was necessary for the lowest
$V$ range 
($18.26\leq V\leq19.26 $) in deriving the Luminosity Function.

We have defined a "central cluster
region" within a radius of $27.07$ arcmin from the cluster centroid and
a "field region" of $1.12\times 0.23$ square degrees  
to estimate the contamination of background and foreground stars.
An initial determination of the Luminosity and Mass Functions has been obtained
assuming that all the cluster stars lie in 
 the "central cluster region". By applying an iterative procedure we have
 estimated the number of cluster stars
within the "field region"  in order to obtain a more accurate correction 
of the cluster
Luminosity Function for field star contamination. 

Evidence for mass
segregation   and energy equipartition have been found from the
spatial distribution of the stars. Extrapolation of the \citet{king62}
empirical model   
allows us to infer that, while all the  stars with $M\geq1.7~M_\odot$
are inside the "central cluster region",
 a non negligible fraction of lower mass stars lies outside.
In particular, by integrating the spatial distribution within the tidal radius
we are able to estimate the number of cluster stars lying beyond the "central
cluster region"
and within the tidal
radius. Therefore, we are able to estimate a correction to the Mass Function
in order to take into account the cluster dynamical evolution.

 The corrected  Present Day Mass Function  was compared with that of the
Pleiades which is known down to
 the brown dwarf limit ($0.08~M_\odot$).
We found that, in the mass range $0.9\leq M/M_\odot\leq  2.5$,
the Mass Function of \ngc2422 
can be represented by a power law of index $\alpha = 3.09\pm0.08$ (rms)
 comparable with the index $\alpha =2.63 \pm 0.15$ obtained
 for the Pleiades in the
same mass range. The index is also consistent with
the data presented in the  log\,$M$ vs. $\alpha$ plot of \citet{krou02}. 

By taking into account  the 
correction due to the dynamical evolution, we computed   a lower limit of
the total mass of \ngc2422  
as $M_{\rm c,cor}\simeq453 M_\odot$.  This value   
is similar to that  of the Pleiades.
 
A future spectroscopic study and a deep survey on a wider field
of this cluster will allow us to find an
independent membership criterion and to extend the Mass Function
below the mass limit of the present
survey.   

\begin{acknowledgements}
We acknowledge P.B. Stetson for having made available to us his 
software and for useful discussions. This work has been  partially
supported by MIUR.
\end{acknowledgements}
\bibliographystyle{apj} 
\bibliography{H3672}  
 
\end{document}